\documentclass[twocolumn,aps,prl,showpacs]{revtex4-1}
\usepackage{amsmath}
\usepackage{amssymb}
\usepackage[tight,TABTOPCAP]{subfigure}
\usepackage{graphicx}
\usepackage{dcolumn}
\usepackage{bm}
\usepackage{setspace}
\makeatletter

\begin{document}
\title{Effects of a particle-hole asymmetric pseudogap on Bogoliubov quasiparticles}

\author{J. P. F. LeBlanc$^{1,2}$}
\email{leblanc@physics.uoguelph.ca}
\author{J. P. Carbotte$^{3,4}$}
\author{E. J. Nicol$^{1,2}$}%
\affiliation{$^1$Department of Physics, University of Guelph,
Guelph, Ontario N1G 2W1 Canada} 
\affiliation{$^2$Guelph-Waterloo Physics Institute, University of Guelph, Guelph, Ontario N1G 2W1 Canada}
\affiliation{$^3$Department of Physics and Astronomy, McMaster
University, Hamilton, Ontario L8S 4L8 Canada}
\affiliation{$^4$The Canadian Institute for Advanced Research, Toronto, ON M5G 1Z8 Canada}
\date{\today}
\begin{abstract}

We show that 
in the presence of a pseudogap, the spectral function in
 the superconducting state of the underdoped cuprates
 exhibits additional Bogoliubov quasiparticle peaks
at both positive and negative energy which are
revealed by the particle-hole asymmetry of the pseudogapped energy bands.
This provides direct information on the unoccupied band via measurement of the occupied states.
When sufficiently close, these Bogoliubov peaks will appear to merge 
with existing peaks leading to the anomalous observation, seen in 
experiment, that the carrier spectral density broadens with 
reduced temperature in the superconducting state.  Using the
resonating valence bond (RVB) spin liquid model in conjunction with recent 
angle-resolved photoemission spectroscopy (ARPES) data allows for an 
empirical determination of the temperature dependence of the pseudogap 
suggesting that it opens only very gradually below the pseudogap onset
temperature $T^*$. 
\end{abstract}
\maketitle

Over the history of high temperature superconductivity in the cuprates,
many fundamental questions have been posed and some have been answered.
It is now known that the charge carriers form in Cooper pairs\cite{keene:1987},
with a pairing symmetry which is described as spin-singlet\cite{barrett:1990}
$d_{x^2-y^2}$-wave\cite{tsuei:2000} and the mechanism might
be spin fluctuations\cite{carbotte:1999}, although the latter issue is
still a subject of considerable debate. Even the applicability of
standard BCS theory has been questioned although experiments have been
presented which give overwhelming evidence for a BCS description.
One such experiment, which is both relevant to this paper and 
demonstrates the impact of the high $T_c$ field to encourage experimental
innovation and improvements in technique, has been ARPES. In ARPES,
not only has the superconducting energy gap $\Delta_{\rm sc}({\boldsymbol{k}})$
been  determined as a function of momentum ${\boldsymbol{k}}$,\cite{damascelli:2003}
but also the predicted Bogoliubov quasiparticle (BQP) bands $\pm E_{\boldsymbol{k}}
=\pm\sqrt{\epsilon_{\boldsymbol{k}}^2+\Delta_{\rm sc}^2({\boldsymbol{k}})}$
 and BQP amplitudes $u^2_{\boldsymbol{k}}$ and $v^2_{\boldsymbol{k}}$ have
been observed and verified to agree
 with $d$-wave BCS theory.\cite{campuzano:1996,matsui:2003,yang:2008}
While this has provided important advances to our understanding of the cuprates
at optimal and overdoping,
it was quickly noted that for underdoped cuprates, the picture was less clear.
Indeed many properties of the superconducting state appear non-BCS-like
and the normal state harbors a not-yet-understood energy-gap-like feature
termed the ``pseudogap''.\cite{timusk:1999} The proximity to the antiferromagnetic
Mott insulator  suggests strong correlation
effects with possibly some competing order and hence the major questions
in the cuprates revolve around understanding the source of the 
pseudogap and its relation to superconductivity. 
Indeed, interest in these issues extend more broadly to the cold atom field of research where a pseudogap has been seen in a strongly interacting Fermi gas using a momentum resolved radio-frequency spectroscopy as an analogue to ARPES.\cite{gaebler:2010}

At present two general
points of view exist. One is that the pseudogap is simply an image of
the superconducting gap related to the existence of phase incoherent preformed pairs
above $T_c$.\cite{emery:1995} This is a one-gap scenario and  argues for the
pseudogap to open symmetrically about the Fermi surface. 
The second point of view 
treats the pseudogap as a manifestation of competing order with
a second energy scale which, along with the superconducting gap,
presents a two-gap scenario.\cite{chakravarty:2001,zhu:2001} Key to this latter vision is that 
 the pseudogap opens up about a surface
in the Brillouin zone which is different from the Fermi surface. For instance,
in the case of competing magnetic order, the pseudogap should be associated
with the antiferromagnetic Brillouin zone boundary. Regardless
of the details of specific models, the two-gap scenario
suggests that the pseudogap will be particle-hole asymmetric. Consequently,
experimental evidence of symmetry or asymmetry would allow for the
elimination of a number of models and provide a significant advancement to the
field. In this letter, we discuss the effect that a particle-hole
asymmetric pseudogap has on the observation of BQPs and propose that
the anomalous broadening of the spectral function seen in ARPES\cite{shen:2010}
 results
from particle-hole asymmetry.

\begin{figure}
  \centering
  \includegraphics[width=0.9\linewidth]{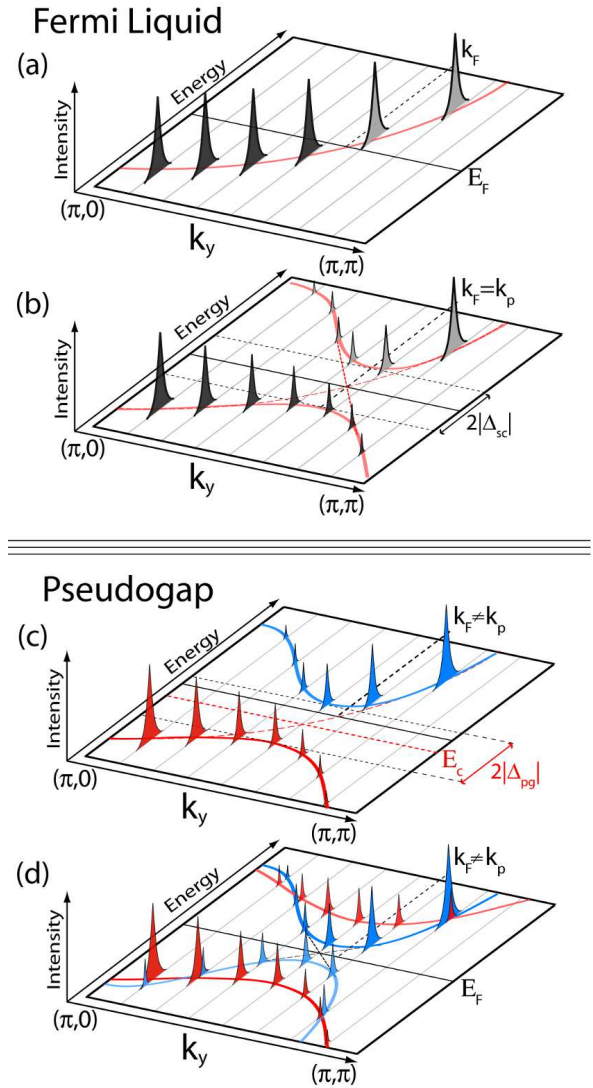}      
  \caption{\label{fig1}(color) Schematic diagram of 
spectral function intensity,
$A({\boldsymbol{k}},\omega)$, 
as a function of energy and momentum for a single Fermi liquid band [(a) and (b)]
and two asymmetric pseudogapped bands [(c) and (d)], as described in the text. 
(b) and
(d) show the formation of BQP bands in the superconducting state.
In the presence of superconductivity, extra spectral peaks in the pseudogapped 
case at fixed ${\boldsymbol{k}}$ 
are revealed that would not be separately resolved
 for a pseudogap opening symmetrically about the Fermi level.}
\end{figure} 

In this work, we focus on ARPES and measurements of the
spectral function $A(\boldsymbol{k},\omega)$. In an ordinary Fermi liquid, 
the spectral function is a simple peak or delta function which tracks the
single-particle energy dispersion $\epsilon_{\boldsymbol{k}}$ as a 
function of energy $\omega$ and momentum $\boldsymbol{k}$ as shown schematically
in Fig.~\ref{fig1}(a). As ARPES only measures the occupied states at zero
temperature due to a Fermi function cutoff at the Fermi level $E_F$,
the peaks above $E_F$ are not detected.  At finite $T$, some information on the bands above $E_F$ can be obtained via analysis of the thermal tails.\cite{matsui:2003} In the presence of superconductivity,
the elementary excitations are the BQPs which mix electron and hole
states. This leads to a gap of $2\Delta_{\rm sc}$ in the dispersion
and introduces two BQP bands  $\pm E_{\boldsymbol{k}}
=\pm\sqrt{\epsilon_{\boldsymbol{k}}^2+\Delta_{\rm sc}^2({\boldsymbol{k}})}$
which show back-bending from the Fermi energy and about the Fermi momentum
$k_F$ which coincides with the position, $k_p$, of the peak in the
back-bending of the occupied states. The two BQP
branches also acquire weighting of 
$u^2_{\boldsymbol{k}}=(1+\epsilon_{\boldsymbol{k}}/E_{\boldsymbol{k}})/2$
and $v^2_{\boldsymbol{k}}=(1-\epsilon_{\boldsymbol{k}}/E_{\boldsymbol{k}})/2$,
as illustrated in Fig.~\ref{fig1}(b). These pictures and their manifestation
in experiment on an overdoped cuprate material were published by
Matsui et al.\cite{matsui:2003} providing a confirmation of the applicability
of $d$-wave BCS theory.
When a pseudogap exists in the normal state, the picture alters and
the energy dispersion will be split into two bands as shown in Fig.~\ref{fig1}(c).
Furthermore, in the presence of a particle-hole asymmetric pseudogap,
$\Delta_{\rm pg}$, the bands do not open about $E_F$ but rather about
some other energy $E_c$ and the back-bending peak position $k_p\ne k_F$,
as shown. In the presence of superconductivity, each band now splits into
two BQP bands positioned symmetrically about $E_F$ [shown as the blue (red)
pair of curves and quasiparticle 
peaks for the upper (lower) band in Fig.~\ref{fig1}(d)].
This gives rise to four BQP peaks as a function of energy, at fixed 
$\boldsymbol{k}$, with two positioned at negative energy and two at
positive energy for $E_F$ taken as 0. Note 
that if the pseudogap opens in a particle-hole
symmetric fashion, the two BQP peaks at negative energy would merge
into one as would the two above $E_F$. It is the combination of
the particle-hole asymmetric pseudogap with the symmetric superconducting
gap which allows the extra hidden BQP peaks to be revealed and displayed
separately.  Therefore, information on the unoccupied band above $E_F$ can now be obtained from analysis of its reflected BQP band on the occupied side, avoiding, in principle, the need for analysis of thermal tails above $E_F$.

To facilitate our discussion, it is necessary to adopt a particular
model. 
 For this purpose we use a model proposed by
Yang, Rice and Zhang \cite{yrz:2006,yrz:2009} who developed an ansatz for the electronic
Green's function based on an RVB spin liquid state. This model has
a particle-hole asymmetric pseudogap and explains a large
amount of anomalous data from the underdoped cuprates.\cite{valenzuela:2007,yang:2010,leblanc:2009,leblanc:2010, carbotte:2010, illes:2009,borne:2010}
Indeed, we will show here
that it is more effective at explaining recent ARPES data\cite{shen:2010} than a
competing model, the $d$-density wave (DDW) model.\cite{chakravarty:2001} Within the ansatz for
the RVB state proposed by Yang et al., the coherent part of
the spectral function is given as 
\begin{equation}
A(\boldsymbol{k},\omega)=\sum_{\alpha=\pm}g_tW^\alpha
[u_\alpha^2\delta(\omega-E_{\rm sc}^\alpha)+v^2_\alpha\delta(\omega+E_{\rm sc}^\alpha)],
\end{equation}
where the energy of the gapped excitations in the superconducting state
are
$E_{\rm sc}^{\alpha}=\sqrt{(E^{ \alpha })^2  +
  \Delta _{\rm sc}^2}$, with Bogoliubov amplitudes
$u_\alpha^2=(1+E^\alpha/E^\alpha_{\rm sc})/2$
and $v_\alpha^2=(1-E^\alpha/E^\alpha_{\rm sc})/2$ 
which are applied to the pseudogapped
bands indexed by $\alpha=\pm$ and given as $E^\pm
=\epsilon_1\pm\sqrt{\epsilon_2^2 +\Delta_{\rm pg}^2}$. 
Here, 
$\epsilon_1=(\xi_{\boldsymbol{k}}-\xi_{\boldsymbol{k}}^0)/2$ and 
$\epsilon_2=(\xi_{\boldsymbol{k}}+\xi_{\boldsymbol{k}}^0)/2$,  
where $\xi_{\boldsymbol{k}}$ is a third nearest-neighbor tight-binding 
dispersion and $\xi_{\boldsymbol{k}}^0$ is that for first nearest-neighbor
which for $\xi_{\boldsymbol{k}}^0=0$  defines the antiferromagnetic
Brillouin zone boundary. $W^\pm$ are weighting factors for
the pseudogapped bands in analogy with the $u$'s and $v$'s and $g_t$ is a 
Gutzwiller factor that reflects a reduction in the coherent part of the
spectral function due to strong correlations. \cite{carbotte:2010}

\begin{figure}
  \centering
 \subfigure{\includegraphics[width=0.9\linewidth]{2by2.eps}\label{fig:1}}    \\            
  \subfigure{\includegraphics[width=0.9\linewidth]{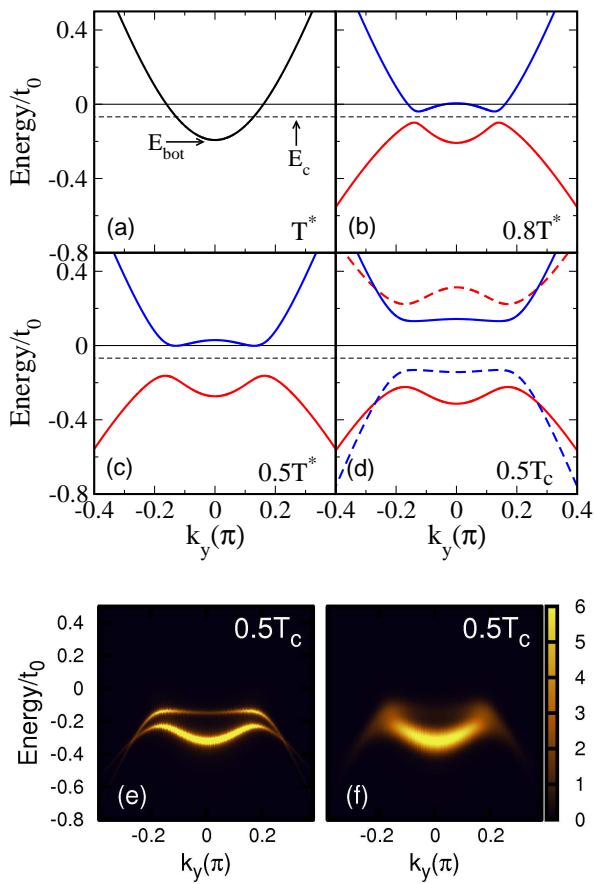}\label{fig:2:b}} 
  \caption{(color) The band dispersions about the point $\boldsymbol{k}=(\pi,0)$ for
a momentum cut along $k_y$. (a) Fermi liquid state at $T=T^*$. The pseudogap
state for (b) $T=0.8T^*$ where the upper band still shows below the Fermi level
and (c) $T=0.5T^*$. (d) The superconducting state shows four BQP bands which
arise from the two particle-hole asymmetric pseudogap bands. (e) gives
the case of (d) now presented as a color map of $I=A(k,\omega)f(\omega)$.
(f) Same as (e) except $I$ is now convoluted with a Gaussian of $\sigma=0.04t_0$.}\label{fig2}
\end{figure} 

In Fig.~\ref{fig2}, we show how the bands change with temperature. 
As in the experiment that we compare to (Ref.~\cite{shen:2010}), the
dispersions are presented as a function of $\boldsymbol{k}=(\pi,k_y)$,
with $k_y$ varying about zero. This is a momentum cut in the antinodal
region of the Brillouin zone where the pseudogap is maximal.
In Fig.~\ref{fig2}(a), the Fermi liquid state at $T=T^*$ gives a single
band which dips below the Fermi level as seen in the experiment. As the
pseudogap develops in (b) and (c), the gap opens about a line below the
Fermi level, breaking particle-hole symmetry. Initially, a double dip
feature appears in the upper band positioned near the Fermi level [Fig.~\ref{fig2}(b)]
as is also seen in experiment\cite{shen:2010} (we do not
find such a feature in the DDW model). 
In the superconducting state shown in 
(d), the secondary BQP bands appear, shown with the dashed curves, which are
mirror reflections about the Fermi energy of the original pseudogapped bands.
An image of the original unoccupied band (solid blue) is now seen on the occupied side (dashed blue).
Fig.~\ref{fig2}(d) is
shown again in (e) as a color map representing the intensity, $I=A({\boldsymbol{k}},\omega)f(\omega)$, where the cutoff due to the Fermi function, $f(\omega)$, is applied
and the quasiparticle weights are included. One sees the two bands at negative
energy clearly separated, but not far apart. To represent instrument
resolution, in (f), we show the convolution of (e) with
a Gaussian of standard deviation $\sigma=0.04t_0$, which would correspond
to roughly 5-15 meV depending on the value of $t_0$. The two bands now appear as one broadened band, particularly around the point
of the back-bending peak, near $k_F$, where the experiments reported anomalous broadening.\cite{shen:2010}
Indeed, as the temperature is lowered and $\Delta_{\rm sc}$ increases, the extra BQP band gains weight and the net result
appears as though there is
an anomalous increase in broadening at low temperature. 

\begin{figure}
  \centering
  \includegraphics[width=0.68\linewidth]{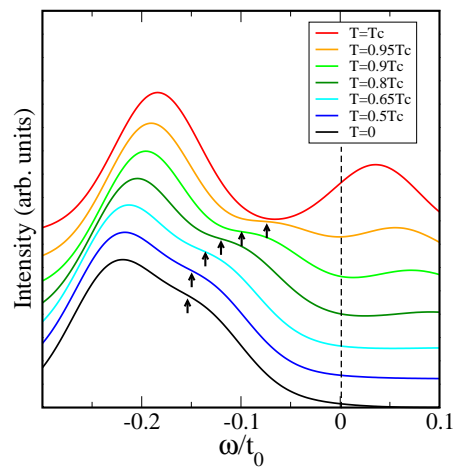}      
  \caption{(color) Intensity $I$ versus energy at $k_F$ [convoluted as in Fig.\ref{fig2}(f)] for various temperatures.  
Each subsequent curve is offset for clarity.  
Arrows track the extra BQP peak emerging at negative energy 
for $T<T_c$.}\label{fig3}
\end{figure}

This is further brought out in Fig.~\ref{fig3} where the intensity
is shown  for several temperatures below $T_c$ at fixed $k_y=k_F$
and for varying $\omega/t_0$.
At $T_c$, the intensity contains two peaks which are not symmetric about $E_F$.  Recent experiments indicate that a minimum in the spectral intensity  occurs at the Fermi level.\cite{shen:2010,hbyang:2010}  One suggestion for this effect might be the existence of regions of spatially inhomogeneous superconductivity which persist above $T_c$.\cite{gomes:2007}
In the superconducting state there is a 
second weaker BQP peak at negative energy which, due to the convolution, 
appears as a shoulder on the main peak.
This shoulder-type feature  (which is traced by the arrows) also exists in the experimental data of Hashimoto et al.\cite{shen:2010} and disappears
above $T_c$ and, while unexplained in the experimental work, it acquires a
natural explanation here as the BQP band arising from the second pseudogap
band at positive energy. 

Much of our discussion to this point has been generic to any model displaying
particle-hole symmetry breaking. Now we address more specifically
the issue of the temperature dependence of the pseudogap, which can be inferred from experiment, and demonstrate that the model used here
is able to explain the distinct qualitative 
features of the data that a competing model,
the DDW, cannot.

\begin{figure}
  \centering
  \includegraphics[width=0.92\linewidth]{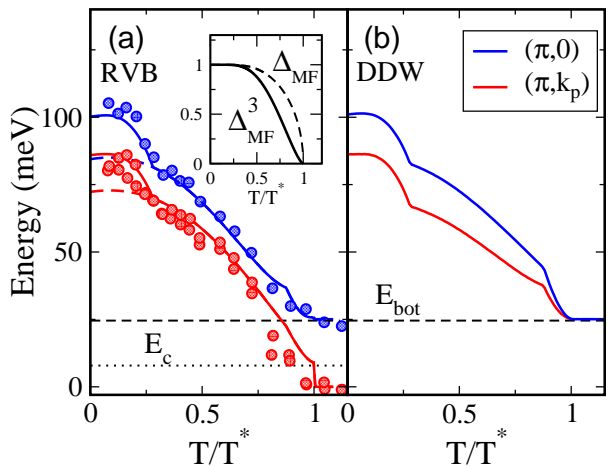}
  \caption{\label{fig4}(color) (a) Temperature-dependent energy of lower band 
at ($\pi,0)$ and ($\pi$,$k_p$),
with and without superconductivity (solid and dashed lines, respectively).
Data from Ref.~\onlinecite{shen:2010}, scaled by $T^*$ in the x-axis, are shown as dots.  
Using  $\Delta_{\rm MF}^3(T)$ [see inset] for the temperature-dependence
of the pseudogap gives a good fit to the data with the RVB model, whereas
$\Delta_{\rm MF}(T)$ does not.
 (b) The DDW model differs from the RVB model and cannot explain the data at $T^*$.}
\end{figure}

In Fig.~\ref{fig4}(a) we plot the energy of the lower band at $(\pi,0)$ and at $(\pi,k_p)$ (the position of the peak in the back-bending) as a function of temperature. 
The curve for ($\pi$,$k_F$) is similar to ($\pi,k_p$). 
To obtain a good fit to the data of Hashimoto et.~al.,\cite{shen:2010} we have adjusted the band structure parameters to fit the antinodal region of the normal state Fermi surface and have used a pseudogap value of 84 meV and a superconducting gap on the Fermi surface of 24 meV.  Along with $t_0=300$ meV, these values are close to those obtained by Yang et.~al.\ \cite{yang:2010} in their consideration of Andreev reflection in an underdoped Bi-based sample of similar $T_c$, which provides support to both of our fits.  In our prior figures, we used a mean field temperature dependence, $\Delta_{MF}(T)$, for both $\Delta_{\rm sc}$ and $\Delta_{\rm pg}$.
The actual temperature dependence of $\Delta_{\rm pg}$ is 
still open to debate. Some argue for the pseudogap
feature in the density of states to fill but not close,\cite{fischer:2007} suggesting a flat
$T$-dependence with a sudden drop at $T^*$.  
With a mean field temperature dependence  for the two gaps (dashed line in the inset) we were not able to agree with the nearly linear temperature dependence observed in the data between $T_c$ and $T^*$.
However,
choosing $\Delta_{\rm pg}(T)$ to have a $\Delta^3_{MF}(T)$ behavior
 we find good agreement with experiment, 
suggesting that the pseudogap may open more gradually in
temperature than previously thought.

In Fig.~\ref{fig4}(b) we 
compare this RVB model and a DDW model. These two models differ
in that the pseudogap opens about the antiferromagnetic Brillouin zone
boundary, $\xi^0_{\boldsymbol{k}}=0$, and hence at $(\pi,0)$ for the DDW and about a surface $\xi_{\boldsymbol{k}}+\xi^0_{\boldsymbol{k}}=0$ for the RVB model which is offset from
the region of $(\pi,0)$. Keeping the bandstructure parameters the same,
along with the temperature dependence of the gaps, we find a qualitative
difference between the two models. The two curves for the DDW model merge to
the same point at $T^*$, ie. the back-bending peak closes to an energy which
is the bottom of the Fermi liquid band,  $E_{\rm bot}$, shown
in Fig.~\ref{fig2}(a) and located at $k_y=0$. The $(\pi,0)$ curve of 
the RVB model also merges to this point but the $k_p$ curve does not
as the back-bending peak in RVB closes at $E_c$ as can be seen in Fig.~\ref{fig2}(b). The lack of separation of the two curves for
$T=T^*$ in the DDW model excludes it as a candidate for the
pseudogap in comparison with the RVB model.

In summary, the anomalous broadening and shoulder
feature seen in ARPES measurements has a natural explanation in a second peak
due to a BQP band in the superconducting state which can only appear  in the
case of particle-hole asymmetry in the pseudogap state.   Further, we find that the pseudogap closes rather gradually with increasing temperature towards $T^*$.  We have also ruled out the DDW model as an alternative competing order as it cannot explain the present ARPES data. 
As a general final comment, the existence of BQPs is fundamental to our understanding of the nature of the many body coherence in the superconducting state.  The observation and measurement of their spectral weight in ARPES was a significant milestone.  In the underdoped cuprates, the BQP peaks may be further split by a 
particle-hole asymmetric pseudogap
leading to a richness in BQP structure which has only been hinted at in recent experiments.
This effect also allows for the unoccupied bands to be studied in the occupied region of the spectral intensity.
  It would be important to find other systems where this phenomenon would reveal itself more clearly.

  The author acknowledge support from the Natural Sciences and
Engineering Research Council of Canada  and the Canadian Institute
for Advanced Research.  

\bibliographystyle{apsrev}
\bibliography{bib}

\end{document}